\documentclass[doublecol]{epl2}
\usepackage{graphics}
\usepackage{graphicx}
\usepackage{epsfig}
\usepackage{amssymb}

\title{Information propagation and collective consensus in blogosphere: a game-theoretical approach}
\shorttitle{Information propagation and collective consensus etc} 
\author{L.-H. Liu \and F. Fu \and L. Wang\thanks{E-mail:\email{longwang@pku.edu.cn}}}
\shortauthor{L.-H. Liu \etal}

\institute{ Center for Systems and Control, College of
Engineering, Peking University, Beijing 100871, China }

\pacs{89.75.Hc}{Networks and genealogical trees}
\pacs{87.23.Ge}{Dynamics of social systems}
\pacs{02.50.Le}{Decision theory and game theory}

\abstract{In this paper, we study the information propagation in
an empirical blogging network by game-theoretical approach. The
blogging network has small-world property and is scale-free.
Individuals in the blogosphere coordinate their decisions
according to their idiosyncratic preferences and the choices of
their neighbors. We find that corresponding to different initial
conditions and weights, the equilibrium frequency of discussions
has a transition from high to low as a result of the common
interest in the topics specified by payoff matrices. Furthermore,
under recommendation, namely, individuals in blogging networks
refer to additional bloggers' resources besides their nearest
neighbors preferentially according to the popularity of the blogs,
the whole blogging network ultrafastly evolves into consensus
state (absorbing state). Our results reflect the dynamic pattern
of information propagation in blogging networks.}

\begin{document}
\maketitle

\section{Introduction}
Blog, which is short for ``web log'', has gained its ground by
online community as a new mechanism for communication in recent
years~\cite{cohen06}. It is often a personal journal maintained on
the Web, which is easily and frequently updated  by the blogger.
In the past few years, blogs are the fastest growing part of the
World Wide Web (WWW). Advanced social technologies lead to change
of the ways of people's thinking and communicating. In the virtual
space of blogs, which is usually referred to as blogosphere,
bloggers present their perception, life experience, ideas, etc, on
their blogs, which is instantly accessible and open to readers
around the world who can comment on the posts of other bloggers.
This creates the eden of free minds and ideas to trigger sparks of
inspiration. Interestingly, the dynamic hierarchy of links and
recommendations generated by blogs creates powerful collaborative
filtering effect to the tremendous information flow travelling
through the blogosphere everyday. Therefore, the blogosphere
provides an extraordinary online laboratory to analyze how trends,
ideas and information travel through social communities. Further
investigation into the blogosphere will help understand not only
the dynamic pattern of information propagation in such ecosystem,
but also the collective behavior taking place on such social
networks.

Much previous research investigating the phenomenon of information
propagation on networks has been done by adopting classic
Susceptible-Infected-Removed (SIR) model in
epidemiology~\cite{Moreno04a,Moreno04b,Huang06,Boccaletti06}.
Thinking about a rumor spreading on social networks: first, an
ignorant (I) acquires the information from her/his neighbors and
becomes a spreader (S). Finally, she/he loses interest about the
information and no longer spreads it again, and becomes a stifler
(R). Accordingly, SIR models information propagation in which the
stiflers are never again susceptible to the information --- just
like conferring lifetime immunity. Yet, SIRS models the situation
in which a stifler eventually becomes susceptible again. For
instance, in blogosphere, the SIRS model can be interpreted as
follows~\cite{Gruhl}: a blogger who has not yet written about a
topic is exposed to the topic by reading the blog of a friend.
She/he decides to write about the topic, becoming infected. The
topic may then spread to readers of her/his blog. Later, she/he
may revisit the topic from a different perspective and write about
it again. Thus the life cycle of the topics is analogous to the
diseases'.

In the realm of sociology, extensive study of the diffusion of
innovation in social networks has been conducted by examining the
power of world of mouth in innovations
diffusion~\cite{Gruhl,Young02,Morris00}. Generally speaking, there
are two fundamental models describing the process by which nodes
in networks adopt new ideas or innovations: threshold models and
cascade models (see ref.~\cite{Gruhl} and references therein).

In this paper, we empirically investigate the dynamic pattern of
information propagation in blogosphere from game-theoretical
perspective. Individuals in the blogosphere coordinate their
decisions according to their idiosyncratic preferences and the
choices of their neighbors. Assume that individuals have two
choices, $A$ and $B$. The payoff of an individual choosing $A$ or
$B$ is composed of two components: an individual  and a social
component. The individual part comes from one's preference
irrespective of others' choices in the network while the social
component of payoff results from the mutual (reciprocal) choices
of individual neighbors. As a result of such a coordination game,
one adapts her/his strategy by imitation at each time step, that
is, follows the more successful strategy measured by the payoff.
Analogous to replicator dynamics, after generations and
generations, the system arrives at equilibrium state. The dynamic
behaviour of such process of information propagation is affected
significantly by the blogging network structure. Thus we study the
dynamics of information propagation empirically on a blogging
network which is collected by our WWW robot. The dynamic pattern
of information propagation as a result of the common interest in
the topics specified by different payoff matrices is also
observed.

The remainder of this paper is organized as follows. Sec.~II deals
with the blogging network data and explains the model we adopt to
study the information propagation in blogosphere, and Sec.~III
gives out the results and makes some explanations. Conclusions are
made in Sec.~IV.

\section{The blogging network and the model}
\begin{figure}
\includegraphics[width=8cm]{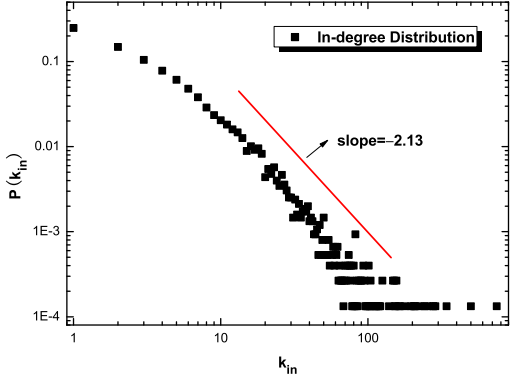}
 \caption{(Color online) The in-degree distribution $P(k_{in})$
  obeys a power-law $P(k_{in})\sim k_{in}^{-\gamma_{in}}$
  with $\gamma_{in}=2.13\pm 0.66$.
  The line's slope is $-2.13$ for comparison with the distribution.} \label{fig.1}
\end{figure}

\begin{figure}
\includegraphics[width=8cm]{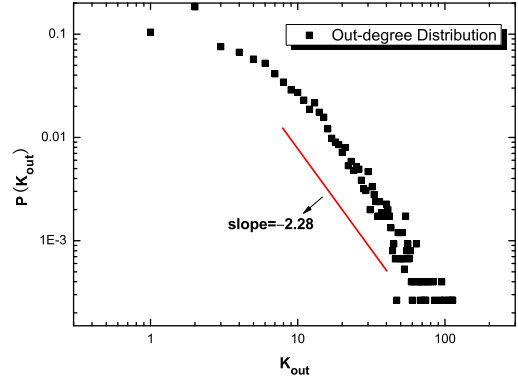}
 \caption{(Color online) The out-degree distribution $P(k_{out})$
  follows a power-law $P(k_{out})\sim k_{out}^{-\gamma_{out}}$
  with $\gamma_{out}=2.28\pm 0.096$.
  The slope of the straight line is $-2.28$ for comparison with the distribution.} \label{fig.2}
\end{figure}

\begin{figure}
\includegraphics[width=8cm]{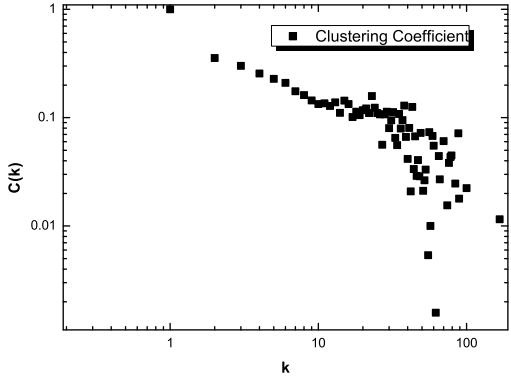}
 \caption{(Color online) The plot of degree-dependent clustering coefficient $C(k)$ versus degree $k$ in undirected blogging network.} \label{fig.3}
\end{figure}

Since the global blogosphere has more than 20 million blogs, we
focused in our preliminary investigation on its sub-community
--- the Chinese blogs hosted by Sina. We viewed this
sub-blogosphere as a close world, i.e., the links outgoing of the
Sina blog community were omitted. We obtained a directed and
connected blogging network consisting of 7520 blogs' sites which
was crawled down by our WWW robot. In fig.~\ref{fig.1} and
fig.~\ref{fig.2}, we report the in- and out-degree distributions
of the directed blogging network. It is found that both in- and
out-degree distributions obey power-law forms where $P(k_{in})\sim
k_{in}^{-\gamma_{in}}$ with $\gamma_{in}=2.13\pm 0.66$,
$P(k_{out})\sim k_{out}^{-\gamma_{out}}$ with
$\gamma_{out}=2.28\pm 0.096$. The average degree of our blogging
network $\langle k_{in}\rangle=\langle k_{out}\rangle=8.42$. We
noticed that about $18.4\%$ of the blogs have no outgoing links to
other blogs, but the in-degree of each vertex in the blogging
network is at least one since our blogging network was crawled
along the directed links. The fraction of reciprocal edges
(symmetric links) is about $31\%$. The degree-dependent clustering
coefficient $C(k)$ is averaged over the clustering coefficient of
all degree $k$ nodes. In fig.~\ref{fig.3}, we can see that for the
undirected blogging network, it is hard to declare a clear power
law in our case. Nevertheless, the nonflat clustering coefficient
distributions shown in the figure suggests that the dependency of
$C(k)$ on $k$ is nontrivial, and thus exhibits some degree of
hierarchy in the network. Besides, the average clustering
coefficient of the undirected blogging network is $0.46$. The
average shortest path length $\langle l \rangle = 4.28$.
Consequently, our blogging network has small-world property and is
scale-free. A detailed study of the structure of the blogging
network is presented in ref.~\cite{Fu_blog}.

Let us introduce the game-theoretical model by which we study the
information propagation on the empirical blogging network. The
social network can be represented by a directed graph $\mathbf{G}$
consisting of a vertex set $V$ and an edge set $E$. Each vertex
$i$ represents a blogger $i$ (her blogs represents herself) in the
blogosphere. A directed edge $e(i,j)$ from $i$ to $j$ indicates
that $j$'s actions influence $i$'s actions. Denote $\Gamma_i$ as
the neighbor set of vertices to which node $i$'s outgoing edges
connect. At each time step, each individual has two choices: $A$
and $B$ corresponding to ``not to discuss the topic (No)'' and
``to write something on the topic (Yes)'' respectively. Let $x_i$
represent individual $i$'s state ($A$ or $B$). For convenience,
the states are denoted by the two-dimensional unit vectors,
\begin{equation}
A=\left(\begin{array}{c}
  1 \\
  0 \\
\end{array}\right )\quad \mbox{and}\quad B=\left(\begin{array}{c}
  0 \\
  1 \\
\end{array}\right )
\end{equation}
The individual's choice depends upon the payoff resulting from
one's idiosyncratic preference and social influence (trend).
Therefore, the payoff of an individual choosing $A$ or $B$ is
composed of two components: an individual and a social component.
The individual part $f_i(x_i)$ of the payoff results from the
intrinsic preference for $A$ or $B$ irrespective of others. The
social component of the payoff is dependent on the externalities
created by the choices of one's neighbors $\Gamma_i$. The social
payoff is supposed to take the form $\sum_{j\in
\Gamma_i}x_i^TMx_j$, where the sum is summed over all $i$'s
outgoing linked neighbors $\Gamma_i$. The payoff matrix $M$ for
the two strategies $A$ and $B$ (the choices $A$ and $B$ can be
interpreted as) is:
\begin{equation}
\begin{array}{ccc}
& A & B \\
A & a & b\\
B & c & d
\end{array}
\label{payoffmatrix}
\end{equation}
where $a>c$ and $d>b$. This is a coordination game where
individuals should choose an identical action, whatever it is, to
receive high payoff. Hence matching the partner's choice is better
off than not matching ($a>c$ and $d>b$). For simplicity, and
without loss of the feature of the coordination game, we set
$b=c=0$ and $d=1-a$ with $0<a<1$. Thus the rescaled payoff matrix
is tuned by a single parameter $a$. The payoff $P_i$ of individual
$i$ is:
\begin{equation}
P_i=(1-w)f_i(x_i)+w\sum_{j\in \Gamma_i}x_i^TMx_j
\end{equation}
where the weight $w\in(0,1)$ indicates the balance between
individual and social payoff. Here we use the strategy update rule
similar to imitation. In any one time step, individual $i$ adopts
the choice $A$ with probability proportional to the total payoff
of her/him and her/his neighbors choosing $A$:

\begin{equation}
W_{x_i\leftarrow A}=\frac{\sum_{j\in
\mathcal{S}_i^A}P_j}{\sum_{j\in\{i\cup \Gamma_i\}}P_j}
\label{update}
\end{equation}
where
$\mathcal{S}_i^A=\{k|k\in\{i\cup\Gamma_i\}\,\,\mbox{and}\,\,x_k=A
\}$. Otherwise, individual $i$ adopts $B$ with probability
$1-W_{x_i\leftarrow A}$. This update rule is in essential
``following the crowd'' in which the individuals are influenced by
their neighbors and learn from local payoff information of their
nearest neighbors. Within this imitation circumstance, the
individual tends to keep up with the social trend based upon the
payoff information gathered from local neighbors.

\section{Results and discussions}
We consider the information propagation on the blogging network
when some bloggers are initially selected at random as seeds for
discussing some certain specified topic in their blogs. All the
bloggers are assumed to be identical in their interests and
preferences, thus the individual part of payoff function
$f_i(x_i)$ is identical for all $i$. For simplicity, we set
$f(A)=0.4$ and $f(B)=0.5$ in our simulations (the same magnitude
as $0<a<1$). In this situation, bloggers preferentially discuss
the topic in their blogs, hence we can examine the world-of-mouth
effect in blogosphere empirically. In addition, all individuals
are influenced by their outgoing linked neighbors. Individual
$i$'s social payoff is summed over her/his all outgoing edges in
which she/he compares her/his choice with her/his friends' and
obtains payoff according to the payoff matrix of
eq.~(\ref{payoffmatrix}). The synchronous updating rule is adapted
here. At each time step, each blogger updates her/his decision
whether to discuss or not according to eq.~(\ref{update}). All
bloggers in the blogosphere are assumed to coordinate their
choices to their friends (whose blogs the outgoing links are
connected to), because conformity with friends in choices leads to
the solid basis to communicate and enjoy the fun of the topics.
Equilibrium frequencies of discussions were obtained by average
over $100$ generations after a transient time of $5000$
generations. The evolution of the frequency of discussions as a
function of $a$ and $w$ has been computed corresponding to
different initial conditions. Furthermore, each data point results
from averaging over 10 runs for the same parameters.

\begin{figure*}
\begin{center}
\includegraphics[width=16.5cm]{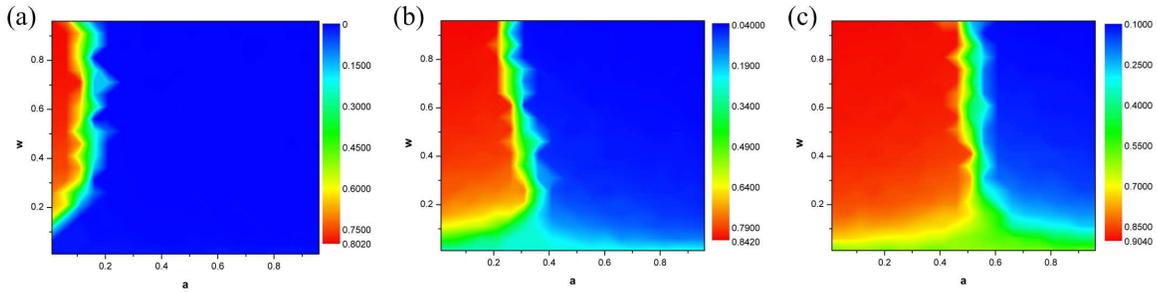}
 \caption{(Color online) The frequency of discussions as a function of the parameter space $(a,w)$. Panel a), b), c) correspond to the initial conditions $1\%$, $21\%$, $51\%$ respectively.} \label{fig.4}
 \end{center}
\end{figure*}

\begin{figure}
\includegraphics[width=8cm]{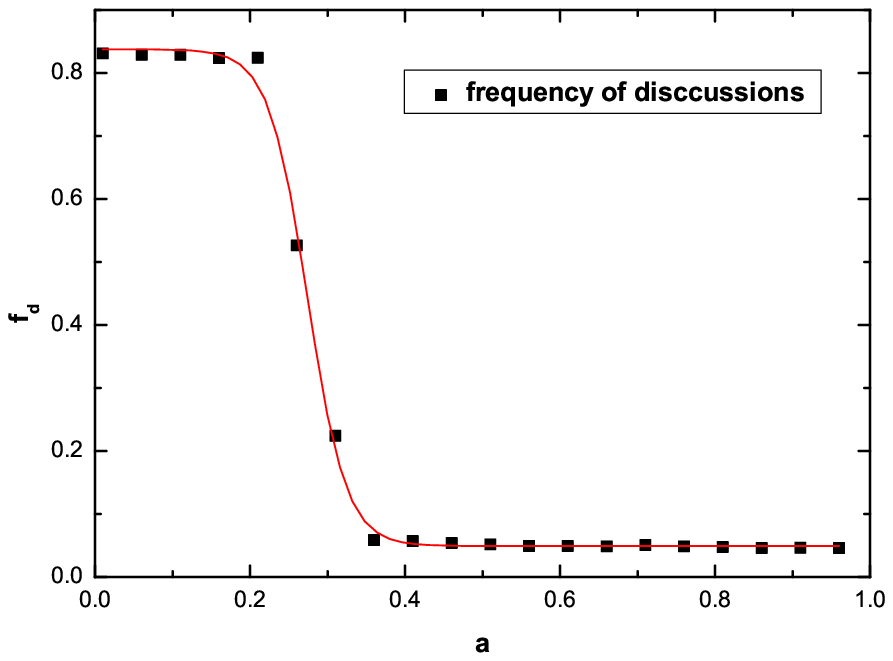}
 \caption{(Color online) The frequency of discussions as a function of $a$ corresponding to $w=0.66$ and initial condition $21\%$.} \label{fig.5}
\end{figure}

\begin{figure*}
\begin{center}
\includegraphics[width=16.5cm]{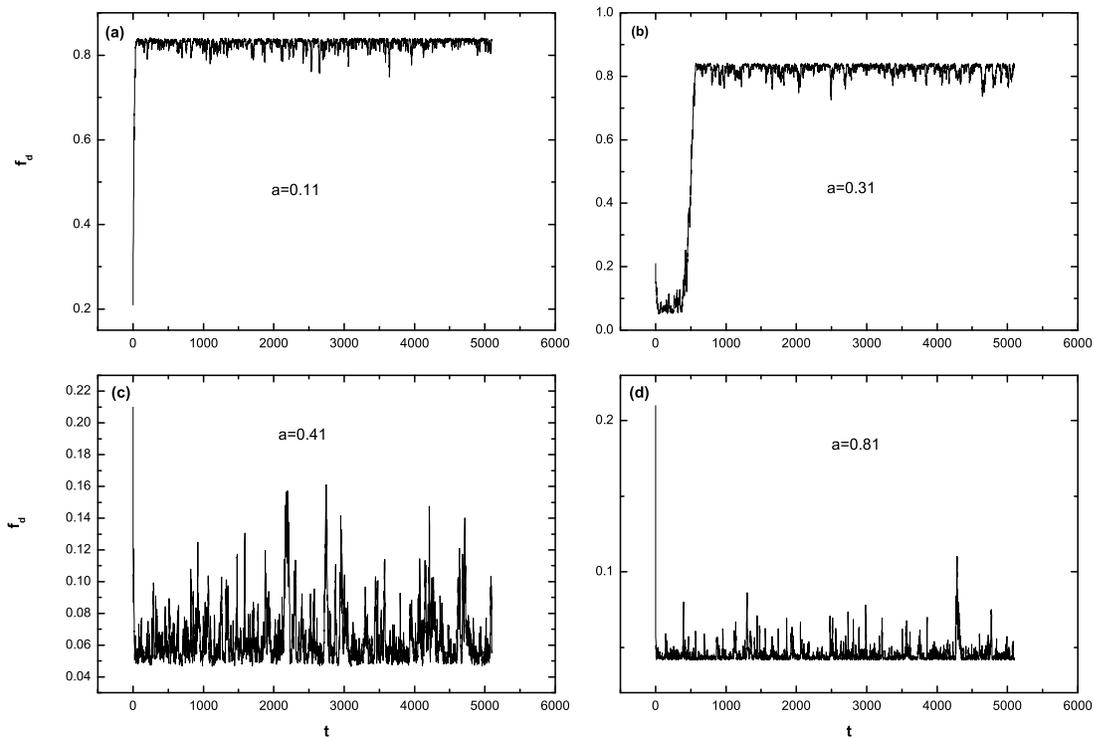}
 \caption{(Color online) Panel a), b), c) and d) show the evolution of discussions corresponding to $a=0.11, 0.31, 41, 0.81$ respectively. The weight $w$ is fixed as $0.66$ and $21\%$ of the bloggers are initially selected as seeds. } \label{fig.6}
 \end{center}
\end{figure*}

We present the results of equilibrium frequency of discussions as
a function of the parameter space $(a,w)$ corresponding to
different initial conditions in fig.~\ref{fig.4}. The density of
discussions is indicated by the color bar. It is found that the
initial fraction of discussers affects the equilibrium results
quantitatively. In fig.~\ref{fig.4}, for a), b), c) panels
respectively, there is a clear transition from high to low for
fixed $w$ when $a$ is increased from 0 to 1. As aforementioned,
the payoff matrix element $a$ indicates the common interest in the
topic travelling in the blogosphere, that is, when $a$ is near
zero, bloggers in the blogosphere show high interest in the topic
and would like to write something about the topic; while $a$ is
near one, it means that people lose interest in the topic, and
reluctant to discuss. Besides, the weight $w$ also plays a role in
equilibrium results. When $w$ approaches to zero, namely,
individuals neglect the social influence and only depend upon
individual preference to discuss or not. While $w$ tends to one,
individuals are completely influenced by their friends regardless
of their own idiosyncratic preferences. Otherwise, for
intermediate $w$, i.e., the choices are balanced between their
individual preference and social influence, the ``self-organized''
bloggers perform in a collective way that without a center control
most of the individuals in the blogosphere change conformably from
frequently discussing the topic to losing interest in the topic as
$a$ increases from zero to one. Moreover, the critical value $a_c$
of $a$ at which the frequency of discussions transits from high to
low is affected by the initial fraction of discussers. It is
observed that for intermediate weight $w$, the critical values of
$a_c$ are around $0.1$, $0.3$, and $0.5$ corresponding to the
initial fractions of discussers $1\%$, $21\%$, and $51\%$
respectively. Thus although the initial condition influences the
equilibrium result, different initial conditions do not change the
equilibrium results qualitatively. In other words, for fixed
weight $w$ and certain initial condition, the density of
discussions has a clear transition from high to low when $a$
increases from $0$ to $1$. We show the frequency of discussions
$f_d$ vs $a$ for $w=0.66$ and $21\%$ of initial discussions in
fig.~\ref{fig.5}. The frequency of discussions decreases from
around $84\%$ to $4\%$ with increasing $a$. The transition happens
around $a_c=0.3$. Interestingly, we find that there are about
$18.4\%$ bloggers have no outgoing links at all. Hence, their
states keep invariant because their decisions are not affected by
neighbors. Accordingly, the whole blogosphere can never evolve
into absorbing states in which all individuals make the same
choice A or B. The typical evolution of frequency of discussions
with respect to time corresponding to different $a$ with $w=0.66$,
initial condition $21\%$ is shown in fig.~\ref{fig.6}. With
$a=0.11$, the blogosphere quickly evolves into the truncated
equilibrium state where the frequency of discussion often drops
down and recovers to previous level after a while. Near the
critical value of $a_c$ with $a=0.31$, the frequency of
discussions is decreased at first. Yet, it strives to achieve the
high level very soon and is retained small fluctuations around the
equilibrium state. When $a$ is increased to $0.41$, the frequency
of discussions descends quickly, and then oscillates around the
equilibrium state. With $a=0.81$, the frequency of discussions
fluctuates with some ``spikes''---occasionally, it suddenly
erupted from $4\%$ to $8\%$. Therefore, these results shown in
fig.~\ref{fig.6} can to some extent reflect the dynamic pattern of
information propagation in the real blogosphere. When most
bloggers show great enthusiasm in the topic (when $a$ is near
zero), they extensively discuss the topic in their blogs. For
example, the fraction of the bloggers talking about ``Microsoft''
in computer community is sustained at high value, even though
there are often small fluctuations around the equilibrium state.
And yet, when all individuals have low interest in discussing the
topic (when $a$ is near one), they are reluctant to mention it in
their blogs. For instance, the discussion of ``influenza'', is
rare at non-influenza season, but bursts out in influenza season.
Consequently, being consistent with the dynamic pattern of
information propagation in real world, our results demonstrate
that the frequency of discussions has a transition from high to
low due to the common interest specified by the payoff matrix for
different weights $w$ and initial conditions.

In order to investigate the role of recommendation to information
propagation in blogosphere, we consider a modified model based
upon the above one. In blogging community, the system often
recommends some recent posts on its main page. Thus the
recommended posts are the information resources which are
noticeable for the bloggers to acquire. In addition, when bloggers
surf in the blogosphere, the probability a blog being visited is
proportional to its in-degree. Therefore, for simplicity, we
assume that besides the neighbors to which the blog outgoing links
connect, each blogger refers to additional $K$ blogs which are
chosen with probability proportional to their in-degrees, i.e.,
the probability $p_{ij}$ that individual $i$ chooses $j$'s blog
($j \nsubseteq \Gamma_i$) as information reference is,
\begin{equation}
p_{ij}=k_j^{in}/\sum_{l}k_l^{in}
\end{equation}
Since each blogger independently chooses $K$ blogs according to
the probability proportional to in-degree, the chosen $K$ blogs of
each blogger might be different. All the individuals are
influenced by both their neighbors and the additional $K$ blogs.
Let $A_i$ denote the $K$ blogs individual $i$ chooses. The payoff
$P_i$ of individual $i$ becomes,
\begin{equation}
P_i=(1-w)f_i(x_i)+w\sum_{j\in \{\Gamma_i\cup A_i\}}x_i^TMx_j
\end{equation}
And the according update rule is,

\begin{equation}
W_{x_i\leftarrow A}=\frac{\sum_{j\in
\mathcal{S}_i^A}P_j}{\sum_{j\in\{i\cup \Gamma_i \cup A_i\}}P_j}
\label{update-2}
\end{equation}
where $\mathcal{S}_i^A=\{k|k\in\{i\cup\Gamma_i\cup A_i
\}\,\,\mbox{and}\,\,x_k=A \}$.

\begin{figure}
\includegraphics[width=8cm]{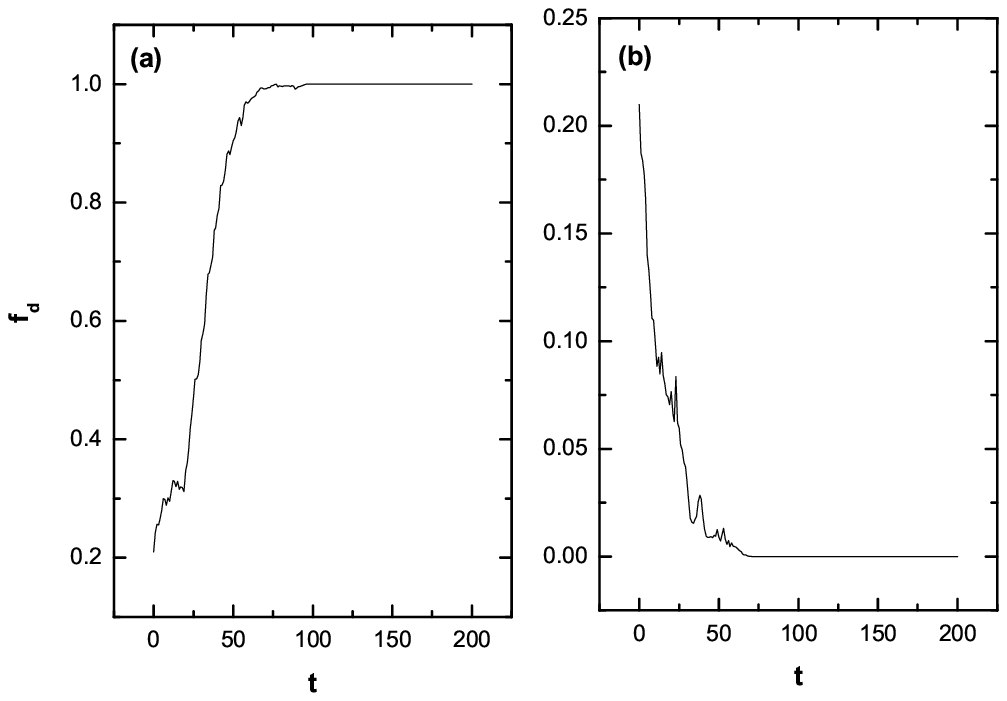}
 \caption{(Color online) The evolution of discussions when individuals choose additional $K$ blogs proportional to the in-degree as references. Left panel a) shows the case with $a=0.11$, and b) with $a=0.81$. The weight $w$ is 0.51, $K=10$, and $21\%$ of the bloggers are initially selected as seeds. } \label{fig.7}
\end{figure}

\begin{figure}
\includegraphics[width=8cm]{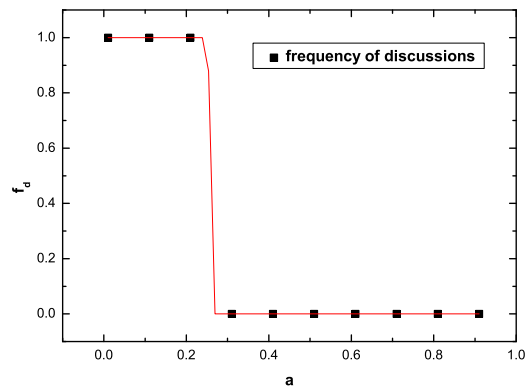}
 \caption{(Color online) The frequency of discussions as a function of $a$ corresponding to $w=0.51, K=10$ and initial condition $21\%$.} \label{fig.8}
\end{figure}
The corresponding results are shown in
figs.~\ref{fig.7},~\ref{fig.8}. Fig.~\ref{fig.7} shows the
evolution of frequency of discussions with $K=10$, $w=0.51$ and
initial condition $21\%$. With $a=0.11$, the whole blogosphere
ultrafastly evolves into absorbing state where all bloggers
discuss the topic in their blogs (see fig.~\ref{fig.7}(a)). While
for $a=0.81$, all individuals choose not to mention the topic at
all (see fig.~\ref{fig.7}(b)). By contrast, without the
recommendation, the whole blogosphere can never evolve into
collective consensus (see
figs.~\ref{fig.4},~\ref{fig.5},~\ref{fig.6} for comparison). In
fig.~\ref{fig.8}, we can see that the frequency of discussions
transits from one to zero as $a$ increases from $0$ to $1$. As a
result, under recommendation, the blogosphere quickly attains the
consensus state in which all bloggers make the same choices A or
B. Herein, the selected (recommended) blogs act as leaders
influencing other bloggers. Dependent upon the local information
$\Gamma_i$ and global information $A_i$, individuals finally
achieve conformity of their choices. Therefore, our result may
shed light on understanding the collective behaviour in the
blogosphere.

\section{Conclusion remarks}
To sum up, we have investigated information propagation on an
empirical social network, the blogging network, by
game-theoretical approach. The blogging network is a good
representative of real social networks which have small-world
property and are scale-free. We found that for different weight
$w$ and initial conditions, the frequency of discussions has a
transition from high to low resulting from the common interest
specified by the payoff matrix. To some extent, our results
reflect the dynamic pattern of information propagation in
blogosphere. Moreover, under the circumstance of recommendation,
the recommended blogs based on their in-degrees act as leaders
influencing other bloggers. Hence, the whole blogosphere evolves
into absorbing states where all bloggers achieve the consensus of
choices. Based upon local information $\Gamma_i$ and limited
global information $A_i$, individual $i$ finally collectively
synchronizes her choice with others. Therefore, our results may
help understand the collective behaviours of bloggers in the
blogosphere.

\acknowledgments The authors are partly supported by National
Natural Science Foundation of China under Grant Nos.10372002 and
60528007, National 973 Program under Grant No.2002CB312200,
National 863 Program under Grant No.2006AA04Z258 and 11-5 project
under Grant No.A2120061303.


\begin{thebibliography}{0}

\bibitem{cohen06}
  \Name{Cohen E. \and Krishnamurthy B.}
  \REVIEW{Computer Networks}{50}{2006}{615}.

\bibitem{Moreno04a}
  \Name{Moreno Y., Nekovee M. \and Pacheco A. F.}
  \REVIEW{Phys. Rev. E}{69}{2004}{066130}.

\bibitem{Moreno04b}
  \Name{Moreno Y., Nekovee M. \and Vespignani A.}
  \REVIEW{Phys. Rev. E}{69}{2004}{055101(R)}.

\bibitem{Huang06}
  \Name{Huang L., Park K., \and Lai Y.-C.}
  \REVIEW{Phys. Rev. E}{73}{2006}{035103(R)}.

\bibitem{Boccaletti06}
  \Name{Boccaletti S. \etal}
  \REVIEW{Physics Reports}{424}{2006}{175}.

\bibitem{Gruhl}
  \Name{Gruhl D. \etal}
  \REVIEW{SIGKDD Explorations}{6}{2004}{43}.

\bibitem{Young02}
  \Name{Young H. P.}
  SFI Working Paper, Paper No. 02-04-018.

\bibitem{Morris00}
  \Name{Morris S.}
  \REVIEW{Review of Economic Studies}{67}{2000}{57}.

\bibitem{Fu_blog}
\Name{Fu F., Liu L.-H., Yang K., \and Wang L.} preprint,
arXiv:math.ST/0607361.

\end{thebibliography}
\end{document}